# A new model for fluid velocity slip on a solid surface

**Jian-Jun SHU**[*], **Ji Bin Melvin TEO** and **Weng Kong CHAN**
School of Mechanical & Aerospace Engineering, Nanyang Technological University,
50 Nanyang Avenue, Singapore 639798.

**ABSTRACT**
A general adsorption model is developed to describe the interactions between near-wall fluid molecules and solid surface. This model serves as a framework for the theoretical modelling of the boundary slip phenomena. Based on this adsorption model, a new general model for the slip velocity of fluids on solid surfaces is introduced. The slip boundary condition at a fluid-solid interface has hitherto been considered separately for gases and liquids. In this paper, we show that the slip velocity in both gases and liquids may originate from dynamical adsorption processes at the interface. A unified analytical model that is valid for both gas-solid and liquid-solid slip boundary conditions is proposed based on surface science theory. The corroboration with experimental data extracted from the literature shows that the proposed model provides an improved prediction compared to existing analytical models for gases at higher shear rates and close agreement for liquid-solid interfaces in general.

## 1. INTRODUCTION

The nature of the boundary condition at a fluid-solid interface has been a long-standing conundrum. The prevailing slip models used are the Maxwell-type collision models for gases and the Navier slip mode, where the slip coefficients may be obtained through the interfacial friction (Bocquet & Barrat 1994, Sokhan & Quirke 2004, Petravic & Harrowell 2007, Kobryn & Kovalenko 2008, Hansen *et al.* 2011). Other interpretations of the fluid-solid interaction involve adsorption concepts (Bhattacharya & Eu 1987, Myong 2004) and thermally activated motion of fluid molecules on a substrate lattice (Ruckenstein & Rajora 1983, Lichter *et al.* 2007).

The state of a fluid molecule upon impact on a surface is governed by interfacial physics and local conditions. When a particle comes into contact with the surface, it has a probability of sticking to the surface or scattering away 'immediately' as shown in Figure 1. Within the kinetic theory framework of the Maxwell slip velocity model, the scattering of particles was classified as specular reflections with no change in particle velocity while the diffuse reflection was akin to the particle being desorbed at the same velocity as the wall. We provide an alternative stochastic interpretation of the molecular conditions at the surface and incorporate the physical details of the fluid-solid dynamics.

---

[*] Correspondence should be addressed to Jian-Jun SHU, mjjshu@ntu.edu.sg

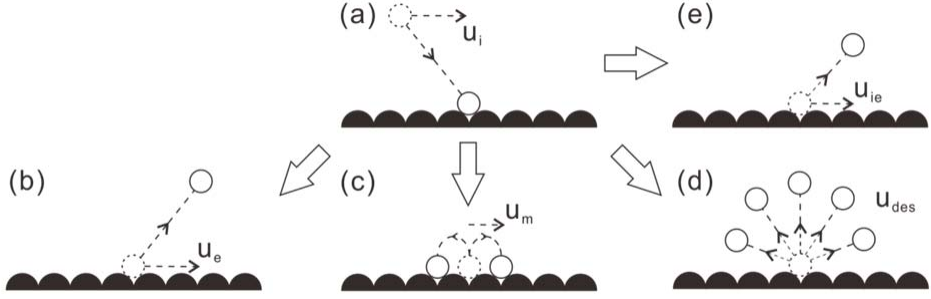

**Figure 1.** Molecular interaction at a fluid-solid interface: (a) incident molecule (b) elastic scattering (c) surface hopping (d) desorption and (e) inelastic scattering.

Sticking, termed as adsorption, occurs by either attraction due to van der Waals (vdW) force or chemical bonding when the particle lands on a vacant binding site on the solid lattice. The former is named physisorption while the latter is known as chemisorption (Ibach 2006). The vdW force, which arises from induced moments in the surrounding atoms as a consequence of charge fluctuation of an atom, dominates at large separation distances $r$, producing an attractive potential with $r^{-6}$ dependence that adheres the particle to the surface. At shorter separations, the vdW force is opposed by the Pauli repulsion which is conventionally assumed to vary as $r^{-12}$. The chemical bonding is much stronger than the vdW force and furthermore is highly directional and site-specific. A transition from activated physisorption to chemisorption is possible with an increase in temperature (Adamson & Gast 2011). In our model, we do not explicitly distinguish between two forms of adsorption with the exception of the range of heat of adsorption being considered.

Within each adsorption site, the thermal motion of the solid and particles results in repeated collisions. An adsorbed fluid particle experiences random forces exerted by the solid particles, which effectively act as a heat-bath. Furthermore, as the gas particle equilibrates with the surface, it also experiences damping forces from the solid that eventually causes it to lose the memory of its initial velocity. This loss in energy is dissipated throughout the solid and the particle's velocity tends towards that of the surface after a characteristic residence time. The dispersion in the velocity caused by the random forces is mediated by the competing effect of damping, which tries to restore the system to its initial state. It is this competition between the opposing effects that gives rise to the equilibrium distribution. Here, we consider temperature that is sufficiently high such that quantum effect can be ignored but low enough for the internal degrees of freedom of particles to be neglected.

The evolution of the tangential velocity $u_a$ of the gas particle (atom or molecule) throughout the duration of interaction with the surface may be described by a Markov process

$$P(u_a, t | u_0, 0)\big|_{t \to 0} = \delta(u_a - u_0) \qquad (1)$$

where $P$ is the transition probability from initial state $u_0$ to $u_a$ at time $t$, $u_0$ refers to the initial velocity at the point of impact and $t$ the residence time of the particle on the surface.

As $t \to \infty$, the probability distribution function (pdf) $p(u_a, t)$ tends toward an equilibrium distribution with mean

$$\langle u_a(t) \rangle = \int_{-\infty}^{\infty} u_r p(u_a, t) du_r = \int_0^t u_a(t') \varphi(t') dt'. \qquad (2)$$

The mean velocity $\langle u_a(t) \rangle$ has the characteristics of a continuous-time random walk in the velocity space with $\varphi(t')$ representing the pdf of the waiting time $t$ between successive velocity jumps and $u_r$ refers to the distribution of the state of the particle at the surface interaction distance $r$.

The tangential velocity of each particle can be modelled classically after the overdamped Ornstein-Uhlenbeck process using the Langevin equation:

$$\frac{du}{dt} = -\gamma u + \eta(t) \qquad (3)$$

where $\gamma$ denotes the damping coefficient and $\eta(t)$ is the noise term that represents the random forces of the solid atoms acting on the gas particle, which conveniently allows us to include the influence of the solid atoms without having to consider the individual motion of each atom. For random forces taking the form of the Gaussian white noise, the mean of the noise term is zero.

$$\langle \eta(t) \rangle = 0. \qquad (4)$$

The initial condition is given by

$$u(0) = u_0 \qquad (5)$$

where $u_0$ is the incident velocity prior to adsorption.

The differential equation in Eq. (3) can be solved together with the initial condition in Eq. (5) to give

$$u(t) = u_0 e^{-t/\tau} + e^{-t/\tau} \int_0^t e^{t'/\tau} \eta(t') dt'. \qquad (6)$$

In Eq. (6) the damping coefficient $\gamma$ in Eq. (3) has been replaced by the reciprocal of the mean sticking time $\tau^{-1}$.

Finally, by averaging over the ensemble, the noise term drops out based on Eq. (4), resulting in the mean tangential velocity expression

$$\langle u(t) \rangle = u_0 e^{-t/\tau}. \qquad (7)$$

Within the scattering regime, the particle reflects specularly (Figure 1(b)) at its original velocity $u_0$ without any exchange of energy with the surface. The sticking time $t_e$ is virtually negligible and can be approximated as (Butt *et al.* 2013)

$$t_e = \frac{2\delta_n}{\langle v_0 \rangle} \approx \frac{2\delta_n}{\sqrt{k_B T/m}} \qquad (8)$$

where $\delta_n$ is the normal penetration distance of the particle into the surface and $\langle v_0 \rangle$ is the normal velocity of the particle which may also be expressed in terms of its thermal energy. Typical room temperature sticking time in this regime for molecular-scale $\delta_n$ is on the order of $10^{-13}$s (Butt *et al.* 2013). Inelastic scattering could occur in individual collisions but this is not expected to affect the equilibrium distribution of the velocity since collisions in which the particles lose energy are cancelled out by those with a gain in energy (Rice & Raw 1974).

The adsorbed particles reside for longer durations of time, during which they interact with the neighbouring solid atoms. For sticking time beyond the mean sticking time $\tau$, most of the gas particles completely thermalise with the surface before being desorbed (Figure 1(d)), emerging with the velocity $u_{des}$ with the tangential component equivalent to that of the wall. Here, we consider a velocity frame of reference relative to the wall such that the desorbed particle leaves with a zero mean relative velocity. This may be termed correspondingly as the fully inelastic regime since the particle retains no trace of its original velocity, having had its initial energy fully dissipated through the solid atoms.

The mean sticking time is given using the Frenkel equation (Frenkel 1924)

$$\tau = \tau_{vib} e^{\Delta H_{ads}/k_B T} \qquad (9)$$

where $\tau_{vib}$ is the inverse of the surface bond vibration frequency and $\Delta H_{ads}$ is the heat of adsorption.

$$\Delta H_{ads} = H_{ads} - H_g \qquad (10)$$

where $H_g$ and $H_{ads}$ are the enthalpies of the gas and adsorbed phase, respectively.

Physisorption takes place typically at around $\Delta H_{ads}$=40kJ/mol with a residence time above $10^{-12}$s (Butt *et al.* 2013). The sticking time for chemisorption has a higher order due to the larger heat of adsorption in the range of $\Delta H_{ads} \geq$40kJ/mol and so the adsorbed particles remain indefinitely on the surface.

An adsorbed particle may also remain mobile in a metastable phase while still being physically bound to the surface if it loses sufficient energy to prevent immediate desorption. In this mobile state (Figure 1(c)), the particle is able to hop to the neighbouring sites with a mean drift velocity $u_m$ before eventually escaping back into the bulk gas or being chemically adsorbed under the right conditions. There is also a probability that an adsorbed particle may escape before reaching thermal equilibrium with the surface (Figure 1(e)), leaving prematurely at the velocity $u_{ie}$ with a portion of its energy dissipated. The partially inelastic regime has an intermediate timescale that ranges between the elastic and mean sticking time. In summary, for sticking time beyond the mean sticking time, $\tau$, particles are desorbed with the same velocity as the wall. However, when $t < \tau$, particles partially retain their initial momentum (when there is flow) and therefore possess a desorption velocity that is non-zero relative to the wall.

## 2. RATE BALANCE EQUATION

We proceed to consider the probability of each interaction type between the fluid and solid particles based on the corresponding rate of the nature of adsorption. The composition of particles departing from the surface comprises those that have undergone either elastic or inelastic interactions. In order to derive the mean condition of particles that leave the surface, the relative rates of sticking and non-sticking events must first be known. The various adsorption processes that take place are dependent on the potential energy landscape of the substrate as well as the energetic conditions of the particles.

The rate of incident particles $R_i$ (Figure 1(a)) may be broken down into the rates of adsorption $R_{ads}$ (Figures 1(c) to 1(e)) and elastic scattering $R_e$ (Figure 1(b)) as follows

$$R_i = R_{ads} + R_e \qquad (11)$$

where $R_{ads}$ and $R_e$ can be expressed in terms of $R_i$ using the sticking probability $p_s$ that represents the fraction of incident particles being adsorbed

$$R_{ads} = p_s R_i \tag{12}$$
$$R_e = (1 - p_s) R_i. \tag{13}$$

The sticking probability is a function of factors such as the surface coverage, temperature, activation energy in the case of activated adsorption and the energy characteristics of the incident particle in non-activated adsorption. In our analysis, $p_s$ is assumed to be a constant parameter, which is valid under the condition of steady-state equilibrium.

Particles that do not scatter upon collision get adsorbed due to energy dissipation during the impact, preventing them from returning to the bulk phase. Among the adsorbed particles, a fraction $p_m$ is loosely trapped but remains mobile in a precursor state while the remaining $1 - p_m$ resides in the potential wells in a physisorbed state, with a possibility of transitioning to the chemisorption state if the temperature rises. At elevated temperature, the precursor state is unfavourable, giving way to direct adsorption followed by desorption.

Though the mobile particles do not possess sufficient momentum in the normal direction to escape, their tangential momentum component allows them to hop from one site in search of another, following which they may desorb after gaining energy either from solid atoms or internally through other degrees of freedom, and get adsorbed at an available site or continue hopping. The rate of adsorption of particles $R_{ads}$ may thus be expressed as

$$R_{ads} = R_{m,1|ads} + R_{s,1|ads} \tag{14}$$

where $R_{m,1|ads}$ represents the rate of adsorbed particles that enter the precursor state and $R_{s,1|ads}$ the rate of adsorbed particles in the stable state.

$$R_{m,1|ads} = p_m R_{ads} \tag{15}$$
$$R_{s,1|ads} = (1 - p_m) R_{ads}. \tag{16}$$

The mobile particles consist of those that remain mobile while others get momentarily adsorbed after landing on an available site. Assuming that all particles in the stable adsorbed state are similar in character and adsorbed particles are eventually desorbed, $R_{m,i|ads}$ during the $i^{th}$ hop is given by the recurring expression:

$$R_{m,i|ads} = R_{m,i+1|ads} + R_{s,i+1|ads} \tag{17}$$

where the subscript $i+1$ indicates the state of the particle after the $i^{th}$ hop.

The adsorbed particles can be segregated into two categories – those that manage to escape while still possessing the parallel momentum with probability $p_e$ and those that undergo desorption with probability $1 - p_e$. The rate of stable adsorption $R_{s,i+1|ads}$ is given by

$$R_{s,i|ads} = R_{ie|s,i|ads} + R_{s|s,i|ads} \tag{18}$$

where $R_{ie|s,i|ads}$ represents the rate of adsorbed particles that escape prematurely and $R_{s|s,i|ads}$ the rate of those that are desorbed after overcoming the energy barrier.

$$R_{ie|s,i|ads} = p_e R_{s,i|ads} \tag{19}$$
$$R_{s|s,i|ads} = (1 - p_e) R_{s,i|ads}. \tag{20}$$

Eqs. (11) to (20) describe the overall rate balance of incident and departure fluxes and can be used in evaluating the mean conditions of the fluid molecules at the surface.

In order to focus on the fundamental essence of adsorption theory in the slip boundary condition, we limit the scope of our study to a steady-state reversible equilibrium process in which the rate of adsorption is matched by the rate of particles being removed from the surface. Furthermore, it is assumed that particles, once stably adsorbed, be it immediately after initial contact with the surface or transitioning from a precursor state, are not physically unique in that they obey similar desorption dynamics. Factors such as lateral interactions between adsorbed particles and a more elaborate form of adsorption like multi-layered adsorption are also ignored.

The foregoing rate balance analysis presented in this paper portrays the complete dynamics of fluid particle interactions with a surface. Using this adsorption framework, we may proceed to derive the mean velocities of the near-wall fluid particles by prescribing the corresponding transport quantities to the respective adsorption states.

## 3. MEAN VELOCITY OF FLUID MOLECULES AT A SOLID SURFACE

The velocity of the particles at the surface can be assessed based on the relative rates of scattering, adsorption and desorption, which can be translated to the probabilities of the respective velocities of each dynamical state.

First, the tangential velocity $u_e$ (Figure 1(b)) of an elastically scattered particle remains unchanged after collision with the surface and is given by

$$u_e = u_i . \tag{21}$$

For particles in the precursor state, their hops can be represented as an asymmetrical random walk. Limiting the motion to one-dimensional uniform jumps and neglecting the influence of other factors such as site vacancy, non-nearest neighbour jumps, correlated jumps *etc.*, the velocity $u_m$ (Figure 1(c)) can be approximated as the drift velocity with the bias being the difference between the rates of hops in the flow direction $R_{m,f}$ and that in the opposite direction $R_{m,b}$.

$$u_m = a(R_{m,f} - R_{m,b}) \tag{22}$$

where $a$ refers to the mean hopping distance.

The velocity of particles that escape in the precursor state is dependent on the duration of adsorption. The dissipation of energy increases with the increasing number of collisions with the solid and relative sliding against adjacent fluid atoms. The partially inelastic desorption velocity $u_{ie}$ (Figure 1(e)) for a particle takes the form

$$u_{ie}(t) = u_i \varphi(t) \tag{23}$$

where $\varphi(t)$ denotes the sticking time distribution.

The fully thermalised particles that have spent an average residence time $\tau_s$ within the wells can be assumed to share the same tangential velocity as the wall upon desorption and therefore emerge with tangential velocity $u_{des}$ (Figure 1(d)) given as

$$u_{des} = 0 \tag{24}$$

where the velocity of each particle is taken in a frame of reference relative to the wall.

Finally, putting together the probabilities of the various adsorption states discussed in Eqs. (11) to (20) and their corresponding velocities in Eqs. (21) to (24), the mean velocity of surface particles has the following expression:

$$\begin{aligned}u_s &= (1-p_s)u_e + p_s p_m u_m + p_s p_e(1-p_m)u_{ie} + p_s(1-p_m)(1-p_e)u_{des} \\ &= (1-p_s)u_e + p_s p_m u_m + p_s p_e(1-p_m)u_{ie}\end{aligned}. \tag{25}$$

Eq. (25) represents the mean velocity of a fluid particle on a solid surface. Whether it is equivalent to the macroscale boundary condition is a recent point of contention (Brenner & Ganesan 2000). Notwithstanding, the interfacial molecular velocity is still relevant in the derivation of a slip velocity on a larger length scale, for instance, by considering a layer of one mean free path thickness as in the treatment of gaseous slip flow. The new boundary condition is applicable to both gas-solid and liquid-solid interfaces although the dominant mechanism of energy or momentum exchange is expected to occur *via* scattering in gases but not in liquids owing to the magnitudes of mean free path. This may offer a plausible reason for the lower slip velocities of liquids, which are mainly due to adsorbed molecules in the precursor state.

## 4. GENERAL SLIP BOUNDARY CONDITION

*Scattering velocity*

The incident velocity of a particle before it arrives at the surface can be linearly approximated by the velocity after its last collision, which in the kinetic theory framework is taken as that from a distance of one mean free path $\lambda$ away.

$$u_i \approx u_s + \lambda \dot{\gamma}_s . \qquad (26)$$

$\dot{\gamma}_s$ denotes the shear rate of the fluid at the surface. Eq. (26) is valid in the range of the low Knudsen numbers $0.001 < Kn < 0.1$ lying in the slip regime (Karniadakis *et al*. 2008). The concept of mean free paths does not readily translate to liquids due to the presence of intermolecular bonds. An alternative parameter that has been suggested as a replacement is the intermolecular bond length.

*Surface diffusion velocity*

In the mobile precursor state, the surface hopping velocity can be modelled after an activated rate process in which case the forward and backward rates take on the Arrhenius form (Ruckenstein & Rajora 1983)

$$R_f = \nu_0 \exp\left(-\frac{E_{a,m} - \Delta E_{shear}}{k_B T}\right), \quad R_b = \nu_0 \exp\left(-\frac{E_{a,m} + \Delta E_{shear}}{k_B T}\right) \qquad (27)$$

where $\nu_0$ is the rate prefactor that has been erroneously identified as the frequency of hopping attempts or vibration frequency in the literature (Ibach 2006), $E_{a,m}$ is the activation energy for surface diffusion and $\Delta E_{shear}$ refers to the change in the potential barrier due to an externally applied shear, which can be approximated as

$$\Delta E_{shear} \approx \frac{1}{2} \mu A_{eff} \, a \dot{\gamma}_s \qquad (28)$$

with $\mu$ being the dynamic viscosity of the fluid and $A_{eff}$ the effective cross-sectional area of a particle under shear. The factor of $\frac{1}{2}$ indicates the lowering of the activation barrier in the direction of shear stress and rising in the opposite direction.

Hence, from Eq. (22), the surface hopping velocity is given by

$$u_m = a(R_{m,f} - R_{m,b}) = u_h \sinh\left(\frac{\dot{\gamma}_s}{\dot{\gamma}_0}\right) \qquad (29)$$

where the free surface diffusion velocity $u_h = v_0 a \exp\left(-\dfrac{E_{a,m}}{k_B T}\right)$ and characteristic shear rate $\dot{\gamma}_0 = \left(\dfrac{\mu A_{eff} a}{2 k_B T}\right)^{-1}$.

The substitution of appropriate values for the parameters reveals that $\dot{\gamma}_0$ is typically on the order of $10^{11} s^{-1}$ for gases and $10^9 s^{-1}$ for liquids, which is at least five orders of magnitude larger than that attainable experimentally for $\dot{\gamma}_s$. Under such conditions, the hyperbolic sine term tends to a first-order function of $\dot{\gamma}_s$. Hence, surface diffusion does not actually contribute to the non-linear dependence on the shear rate in practical situations except at highly exaggerated shear rates such as those investigated in the molecular dynamics (MD) simulations of Wang & Zhao (2011). In their study, it was shown that Eq. (29) provided a fairly good prediction of their MD results at the shear stress values of up to 100MPa although the curve-fitting details were not elaborated. Recalling Eq. (29), in the limit $\dot{\gamma}_s \ll \dot{\gamma}_0$, the surface diffusion velocity can be approximated as

$$u_m \approx u_h \dfrac{\dot{\gamma}_s}{\dot{\gamma}_0}. \tag{30}$$

*Escape velocity*

The escape velocity while the particle is in the precursor state can be expressed by considering its net change in the tangential direction as follows

$$u_{ie} = \sqrt{u_e^2 - \dfrac{2}{m_m}\Delta E} = \sqrt{u_e^2 - 2\mu_u \delta u_e} = \sqrt{u_e(u_e - 2\mu_u \delta)} \tag{31}$$

where $\Delta E$ represents the mean energy loss during the period of sticking in the precursor state, $\mu_u$ the effective friction coefficient, and $\delta$ the average distance traversed. The energy dissipation arises from interactions with the substrate as well as adjacent fluid particles within the bulk flow and can be approximated as velocity-dependent friction based on the relative velocity of the adsorbed particle and the surrounding environment (Krim 2012).

Substituting the velocity expressions in Eqs. (21), (26), (30) and (31) into Eq. (25), the following quadratic expression for the slip velocity can be obtained after rearrangement

$$(1-c_1)u_s^2 - 2[(c_2 + c_1 \lambda)\dot{\gamma}_s - c_1 \mu_u \delta]u_s + [(c_2^2 - c_1 \lambda^2)\dot{\gamma}_s + 2c_1 \mu_u \delta \lambda]\dot{\gamma}_s = 0$$

$$c_1 = p_e^2(1-p_m)^2, \quad c_2 = \dfrac{1-p_s}{p_s}\lambda + \dfrac{p_m u_h}{\dot{\gamma}_0} \tag{32}$$

where coefficients $0 \leq c_1 \leq 1$ and $c_2 \geq 0$ have been introduced.

Solving Eq. (32) for $u_s$ gives

$$u_s = C_1 \dot{\gamma}_s - C_2 \pm \sqrt{\dfrac{C_2(C_1+\lambda)^2}{C_2 + \mu_u \delta}\dot{\gamma}_s^2 - 2C_2(C_1+\lambda)\dot{\gamma}_s + C_2^2}$$

$$C_1 = \dfrac{c_2 + c_1 \lambda}{1 - c_1}, \quad C_2 = \dfrac{c_1 \mu_u \delta}{1 - c_1} \tag{33}$$

Since the slip velocity should cease to exist in the absence of an external field ($u_s|_{\dot{\gamma}_s=0} = 0$), the negative root can be discarded, leaving the final expression

$$u_s = C_1\dot{\gamma}_s - C_2 + \sqrt{\frac{C_2(C_1+\lambda)^2}{C_2 + \mu_u\delta}\dot{\gamma}_s^2 - 2C_2(C_1+\lambda)\dot{\gamma}_s + C_2^2} \qquad (34)$$

where it should be emphasised that the coefficients $C_i(i=1,2) > 0$ are the representative of the interfacial conditions, adsorption probabilities and properties of the media as follows:

$$C_1 = \frac{1}{1-p_e^2(1-p_m)^2}\left[\frac{1-p_s}{p_s}\lambda + \frac{p_m u_h}{\dot{\gamma}_0} + p_e^2(1-p_m)^2\lambda\right]$$
$$C_2 = \frac{p_e^2(1-p_m)^2}{1-p_e^2(1-p_m)^2}\mu_u\delta \qquad (35)$$

Eq. (34) is the main result for this paper and represents a new general slip velocity model for fluid-solid boundary conditions derived based on the theory of interfacial physics, specifically adsorption and desorption processes. The novelty of this model lies in its applicability to both gas and liquid flows, which has thus far been studied independently in analytical models to the best of our knowledge. Furthermore, the slip velocity expression exhibits non-linearity with respect to the wall shear rate which is in accordance with the prediction of experimental measurements where such phenomena have been observed.

## 5. VALIDATION OF SLIP VELOCITY MODEL FOR A GAS-SOLID INTERFACE

While it is remarkable that Maxwell managed to conceive the tangential momentum accommodation coefficient (TMAC) term to describe the effective gas-surface interactions at a point in time when the realm of surface physics was virtually unknown, TMAC reveals little about the physical nature of the inter-molecular interactions and the actual motion of fluid molecules at the interface. Fundamentally, the assumption of elastic scattering represents an ideal situation that disregards the occurrence of inelastic scattering events. The TMAC, which is analogous to the sticking probability, is also not a constant as it should depend on the characteristics of the incident molecule. It is therefore natural that the slip boundary condition should be modelled instead using adsorption-desorption processes. Even so, it has to be acknowledged that the simple form of the linear slip velocity makes it attractive for use in theoretical studies and to date remains a popular area of research for experimentalists.

The Langmuir model marks the first attempt at deriving the slip velocity based on adsorption concepts (Bhattacharya & Eu 1987). However, the simple adsorption model based purely on site vacancy is essentially similar to the Maxwell model with the non-dissociative sticking probability playing a similar role as the TMAC. More importantly, both models are only linearly dependent on the shear rate and thus incapable of explaining experimental results displaying a non-linear trend.

To ensure that our model is physically sound, we compare predictions by our model to experimental and numerical results that are available in the literature. First, the procedure of experimental data extraction is briefly described. Following that, the theoretical curves from both new and existing models are plotted and compared with the extracted data.

*Experimental data for gas-solid interfaces*

The experimental studies selected for comparison involve the measurement of mass flow rates and differential pressure of microchannel gas flows, which can be converted into the slip velocity and the wall shear rate for comparison with the new slip boundary condition. In the low Knudsen slip regime, the velocity profile for the Poiseuille flow through a long, straight microchannel of uniform rectangular cross-section with a low height-width aspect ratio can be simply worked out by solving for the Stokes flow coupled with slip boundary conditions prescribed at the top and bottom walls, giving

$$u = \frac{h^2}{2\mu} \frac{dP}{dx} \left( \frac{y^2}{h^2} - 1 \right) + u_s \tag{36}$$

where $2h$ represents the height of the channel, $\frac{dP}{dx}$ the pressure gradient in the flow direction $x$. The $y$ coordinate is taken to be in the normal direction to the flow with the origin located in the centre of the top and bottom walls.

Subsequently, the mass flow rate can be obtained from the velocity profile and further rearrangement gives the desired slip velocity in terms of the mass flow rate and differential pressure as

$$u_s = \frac{\dot{m}}{\rho w h} - \frac{h^2 \Delta P}{12 \mu l}. \tag{37}$$

The shear rate of the fluid at the wall $y = \pm h$ can similarly be determined by differentiating Eq. (36) with respect to $y$

$$\dot{\gamma}_s = \frac{h \Delta P}{2 \mu l}. \tag{38}$$

After performing the above conversion, the mass flow rate *versus* pressure ratio curves can be transformed into the curves of slip velocity against wall shear rates for the ease of comparison with Eq. (34).

*Comparison with experimental studies for gas-solid interfaces*

First, the mass flow rate measurement data for helium and nitrogen gases in silicon microchannels conducted by Shih *et al.* (1996) are used. The 4mm long channel had a rectangular cross-section 40μm wide and 1.2μm high. The converted wall shear rates were on the order of magnitudes of $10^5$ to $10^6 s^{-1}$, with a slip velocity of up to 0.55ms$^{-1}$. This justifies the use of the linear approximation for the surface diffusion term as indicated in Eq. (30).

In the second study, we use the results of Arkilic *et al.* (1997), who performed the mass flow rate measurements of rarefied helium gas flows in silicon microchannels measuring 52.25μm wide, 1.33μm deep and 7500μm long in the slip regime with a mean outlet Knudsen number of 0.155. The mass flow rates and differential pressures were translated into slip velocities ranging from 0.07 to 0.79ms$^{-1}$ for wall shear rates between $0.25 \times 10^6$ and $1.35 \times 10^6 s^{-1}$.

The third reference is from experimental investigations carried out by Zohar *et al.* (2002) on the flows of helium, argon and nitrogen gas in the silicon nitride coated microchannels of dimensions 40μm in width, 4000μm in length, and 0.53μm in height. The mean Knudsen number for the experiment ranged from 0.118 to 0.384. The range of slip velocities was 0.03 to 0.6ms$^{-1}$ and that for wall shear rates was $0.35 \times 10^6$ to $1.36 \times 10^6 s^{-1}$.

The final set of results being compared is taken from the non-equilibrium MD simulations of Kannam *et al*. (2011) for the Couette flow of argon and methane in 5.78nm tall grapheme nanochannels. The wall shear rates of $0.85\times10^8$ to $1.60\times10^{11}\text{s}^{-1}$ were much higher than the above two experiments due to the computational time scales involved. As such, the slip velocities were also several orders larger, starting from $6\text{ms}^{-1}$ up to a maximum of $8.62\times10^3\text{ms}^{-1}$.

Figures 2(a) to 2(h) show the plots of slip velocity against the wall shear rate from the above sets of extracted data. Also plotted within the same graphs are the best-fit curves using Eq. (34) and that of the existing slip velocity models - jointly represented as a single plot by the simplified linear expression

$$u_s = b\dot{\gamma}_s \tag{39}$$

where $b$ is the slip coefficient. It should be reiterated that all previous analytical models only possess a first-order dependence on the wall shear rate; the so-called second-order models that build upon the Maxwell model to improve its predictions at the moderate Knudsen numbers merely retain the second expansion term of the slip velocity and do not indicate a non-linear relationship with the wall shear rate. For free molecular conditions (*Kn*>10), analytical solutions to the Boltzmann equation for simple geometries can be obtained (Fukui & Kaneko 1988) while MD and direct simulation Monte Carlo method can provide numerical solutions for complex geometries (Huang *et al*. 1997). The modelling of the transition regime (0.1<*Kn*<10), however, remains a problem by virtue of the equal importance of intermolecular and molecule-surface collisions. For the Poiseuille flow, the second derivative of the velocity can therefore be expressed in terms of the wall shear rate while that for the Couette flow vanishes. Thus, Eq. (39) is only fitted to the experimental data for low shear rates where the trend remains linear.

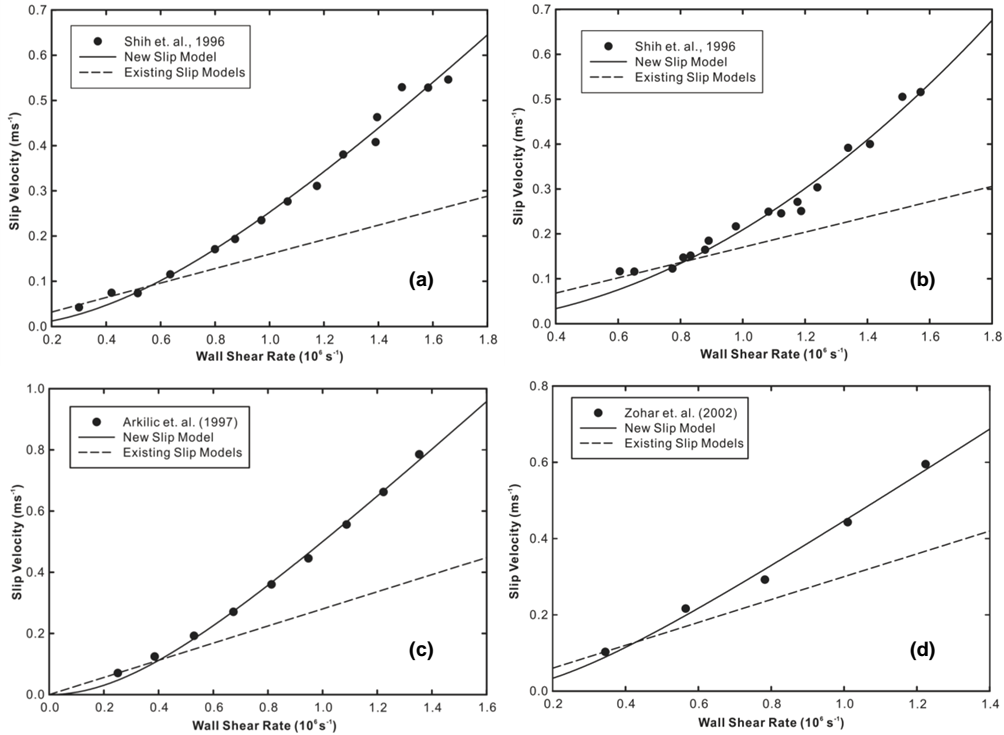

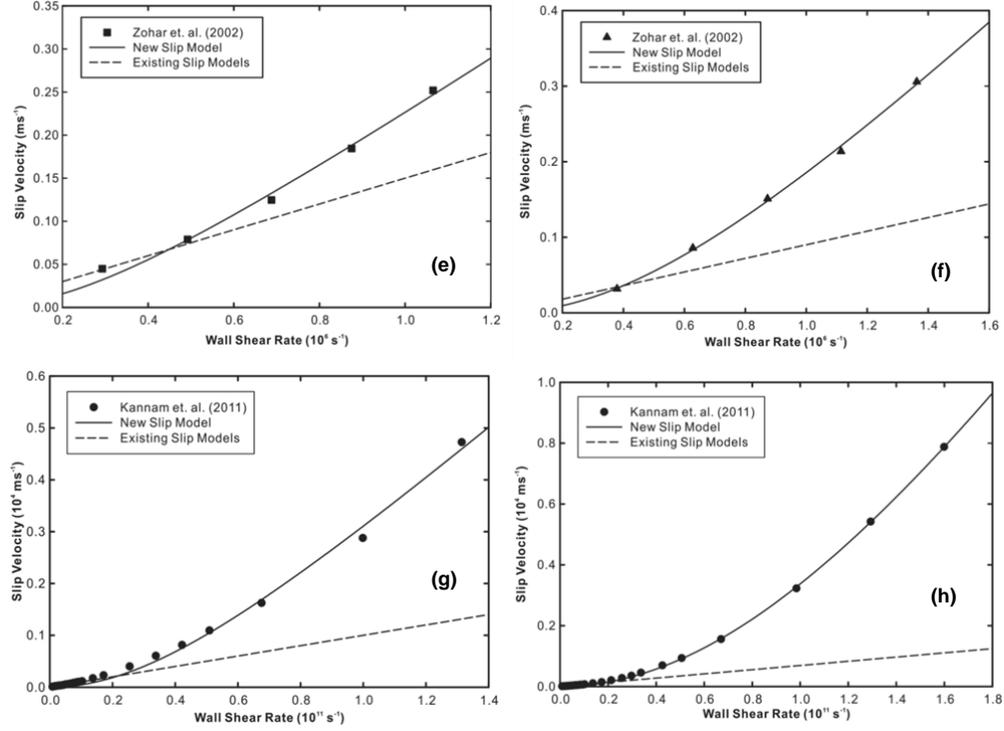

**Figure 2.** Comparison of the new (solid lines) and existing (dashed lines) slip models for gas-solid interfaces using experimental results (symbols) from the literature. The fitting coefficients are summarised in Table 1.

**Table 1.** Details of experiments and values of coefficients in Figure 2

| Authors | Figure | Gas | Surface | $C_1$(m) | $C_2$(ms$^{-1}$) | $b$(m) |
|---|---|---|---|---|---|---|
| Shih *et al*. (1996) | 2(a) | Helium ($Kn$=0.158) | Silicon | $5.77\times10^{-23}$ | $6.02\times10^{-1}$ | $1.6\times10^{-7}$ |
| | 2(b) | Nitrogen ($Kn$=0.054) | | $1.04\times10^{-24}$ | $3.66\times10^{1}$ | $1.7\times10^{-7}$ |
| Arkilic *et al*. (1997) | 2(c) | Helium ($Kn$=0.155) | Silicon | $1.56\times10^{-24}$ | $4.32\times10^{-1}$ | $2.8\times10^{-7}$ |
| Zohar *et al*. (2002) | 2(d) | Helium ($Kn$=0.384) | Silicon nitride | $7.54\times10^{-26}$ | $2.17\times10^{-1}$ | $3.1\times10^{-7}$ |
| | 2(e) | Argon ($Kn$=0.196) | | $1.54\times10^{-28}$ | $1.33\times10^{-1}$ | $1.5\times10^{-7}$ |
| | 2(f) | Nitrogen ($Kn$=0.118) | | $2.62\times10^{-24}$ | $3.31\times10^{-1}$ | $8.9\times10^{-8}$ |
| Kannam *et al*. (2011) | 2(g) | Argon | Graphene | $4.29\times10^{-28}$ | $2.89\times10^{3}$ | $9.0\times10^{-9}$ |
| | 2(h) | Methane | | $8.92\times10^{-24}$ | $2.19\times10^{4}$ | $6.9\times10^{-9}$ |

## 6. VALIDATION OF SLIP VELOCITY MODEL FOR A LIQUID-SOLID INTERFACE

For liquid flows on solid surfaces, a survey of the literature reveals a relative lack of analytical models for slip boundary condition. The majority of the experimental and theoretical studies involving micro- and nano-fluidics mostly employ the Navier slip boundary condition, which oversimplifies the problem due to the use of a constant slip length although the deviation from the model is apparent from the experimental results. Again, we use available results in the literature to demonstrate the conformity of predictions using our model.

*Experimental data for liquid-solid interfaces*

In the drainage force measurement approach, experimental data are normally plotted as a slip length against the nominal flow rate. For a surface force apparatus having identical cylindrical probes, the shear rate can be estimated from the expression

$$\dot{\gamma}_{max} = \sqrt{\frac{27}{128} \frac{R}{h} \frac{v_{peak}}{h}} \qquad (40)$$

where $R$ refers to the radius of the cylinder, $h$ the film thickness and $v_{peak}$ the peak oscillation velocity for sinusoidal vibrations. The details of the derivation of Eq. (40) were provided in the paper by Horn *et al*. (2000). Here, the maximum shear rate is used as a rough estimate since the shear rate varies in the region of measurement due to the curved geometry of the probes.

The rest of the experimental and numerical studies being compared do not require any experimental technique-based conversion apart from the straightforward calculation of slip velocity from the slip length and shear rate values provided using the Navier slip boundary condition.

*Comparison with experimental studies for liquid-solid interfaces*

A total of five experimental and MD studies have been chosen for quantitative comparison with our slip model for liquid-solid interfaces. Given the pronounced non-linearity in most of the experimental data, the prediction of the linear Navier slip velocity is not shown in the graphs. Furthermore, the theoretical surface diffusion model of slip remains linear under experimental conditions as discussed previously and hence does not warrant a comparison with our model predictions.

Two of the selected studies were conducted by Zhu & Granick (2001, 2002), who published a series of experimental findings on the subject of liquid slip with a particular focus on its shear rate dependency. In their experiments, they employed the popular thin film drainage force measurement technique by utilising a surface force apparatus. Slip lengths were inferred from force measurement curves for the assorted liquid films of down to 2nm thickness that were confined between sinusoidally-driven cylindrically-shaped mica probes - each of 2cm radius of curvature and surfactant-coated. The exponential increase in slip length was apparent throughout most of their experimental results.

Another two sets of results are drawn from the studies of Huang *et al*. (2006), who used an imaging technique known as total internal reflection velocimetry to probe the near-wall velocities for the pressure-driven flows of tracer-laden deionised water in a 50µm deep PDMS microchannel at the glass surfaces of different wettabilities. In addition to the higher slip lengths measured for the hydrophobic surface (26 to 57nm and 37 to 96nm respectively for the hydrophilic and hydrophobic surfaces), it was similarly observed that the slip length was not a constant, increasing with an increase in the shear rate.

Ulmanella & Ho (2008) reported the mass flow rate slip measurements of liquid flows of isopropanol and *n*-hexadecane in the micro- and nanochannels of depths between 350nm and 5µm, which were fabricated from glass bonded to a silicon substrate. The roughness of the channel walls was controlled by varying the etchant concentration. This allowed them to produce different surface roughness of 0.5nm and 8.5nm. Slip flow was clearly enhanced in the smoother channels and shown to be independent of the channel heights. Non-linear slip behaviour was also evident in their experimental results.

The last study used is from an MD simulation of *n*-decane in a Couette flow configuration that was carried out by Martini *et al*. (2008). Investigating the role of the wall model used in MD simulations and its resultant influence on the shear rate *versus* slip relationship, they discovered an unbounded increase in slip length with increasing shear-rate for a rigid surface model while that for the flexible surface model remained relatively constant at the wall speeds of up to 1000ms$^{-1}$ for a channel height of 3nm.

Data from the above studies are converted to the values of slip velocity and wall shear rate and reproduced in Figures 3(a) to 3(h), which also contain the least-squares fit using Eq. (34) to provide verification for our newly derived slip velocity model.

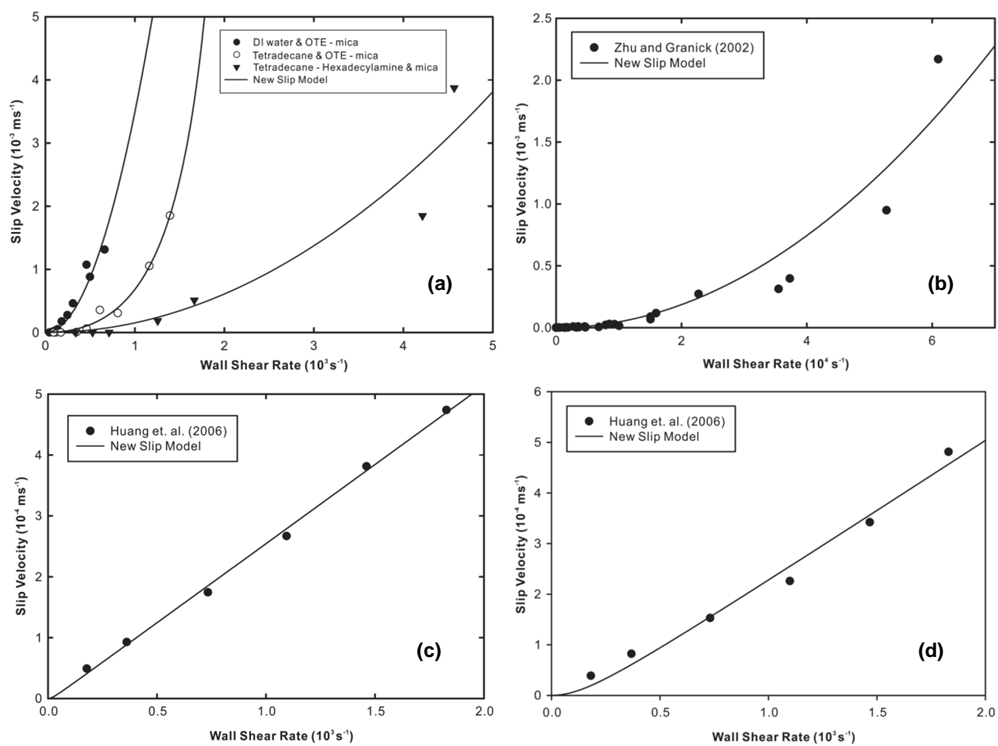

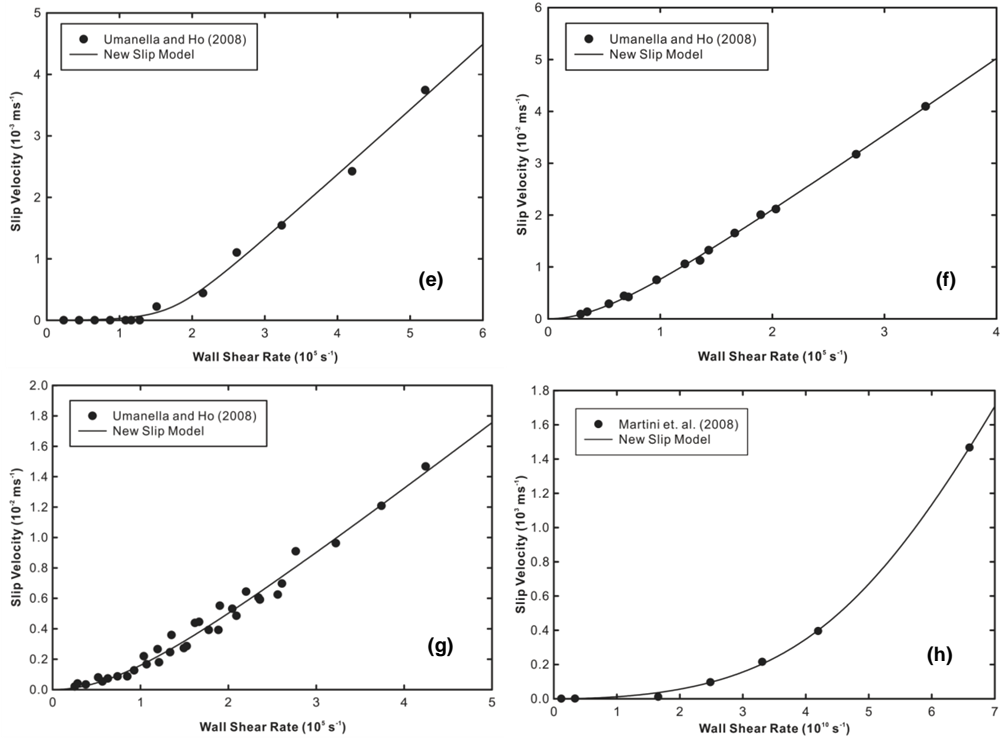

**Figure 3.** Validation of the new slip model (solid lines) for liquid-solid interfaces using experimental results (symbols) from the literature. The fitting coefficients are summarised in Table 2.

**Table 2.** Details of experiments and values of coefficients in Figure 3

| Authors | Figure | Liquid | Surface | $C_1$(m) | $C_2$(ms$^{-1}$) |
|---|---|---|---|---|---|
| Zhu & Granick (2001) | 3(a) | DI water | Mica+OTE | 2.59×10$^{-7}$ | 3.53×10$^{1}$ |
|  | 3(a) | Tetradecane | Mica+OTE | 3.83×10$^{-10}$ | 7.39×10$^{0}$ |
|  | 3(a) | Tetradecane+HDA | Mica | 2.38×10$^{-10}$ | 1.74×10$^{1}$ |
| Zhu & Granick (2002) | 3(b) | DI water | Mica+PVA | 2.55×10$^{-11}$ | 1.19×10$^{3}$ |
| Huang et al. (2006) | 3(c) | DI water | PDMS (hydrophilic) | 9.23×10$^{-27}$ | 5.60×10$^{-6}$ |
|  | 3(d) |  | PDMS (hydrophobic) | 7.16×10$^{-22}$ | 5.62×10$^{-5}$ |
| Ulmanella & Ho (2008) | 3(e) | Hexadecane | Glass/silicon (rough) | 5.20×10$^{-9}$ | 9.93×10$^{-4}$ |
|  | 3(f) | Hexadecane | Glass/silicon (smooth) | 4.29×10$^{-28}$ | 1.10×10$^{-4}$ |
|  | 3(g) | Isopropanol | Glass/silicon | 2.90×10$^{-26}$ | 5.45×10$^{-5}$ |
| Martini et al. (2008) | 3(h) | n-decane | Rigid wall | 3.55×10$^{-8}$ | 2.27×10$^{3}$ |

## 7. DISCUSSION

It is evident from the slip velocity curves for both gases and liquids in Figures 2 and 3 that the experimental data exhibit significant non-linearity at elevated wall shear rates which the existing slip boundary condition models fail to predict. By fitting Eq. (39) to the data lying within the low shear rate regime using constant slip coefficients, it is found that the analytical curves rapidly deviate from the experimental results as the wall shear rate increases. In contrast, our new model matches the experimental results for gas-solid slip velocity more closely at higher shear rates, although a slight deviation is observed when the shear rate is low. The model also shows good agreement for the liquid-solid slip velocity except for discrepancies at high shear rates in Figures 3(a) and 3(b).

The poor agreement of existing theoretical models with the experimental results for gaseous slip as seen in Figures 2(a) to 2(h) can be attributed to the simple scattering law adopted in the kinetic-theory based models, which assumes a constant TMAC. Similarly, the elementary adsorption rule applied in the Langmuir approach corresponds to a constant sticking probability. Conversely, good agreement between our new model and the experimental results supports the idea of an adsorption-desorption based mechanism of fluid slip. This suggests that near-wall particles are not only limited to pure elastic and diffuse collisions but also various adsorption processes that transpire after impact, among which includes the dissipation of energy during escape from the mobile phase that contributes to the non-linear dependence on shear rate. The non-linear dependence on shear rate is dependent on the value of the coefficient $C_2$ in Eq. (34), which reverts to a linear function of shear rate when $C_2$ is zero. Physically, $C_2$ represents the inelastic contribution of the trapping phase relative to the other adsorption states. As seen in the experiment results of Zohar *et al.* (2002) in Figures 2(d) to 2(f) for helium, argon and nitrogen, $C_2$ increases as the non-linearity becomes more pronounced. In descending order, the values of $C_2$ are $0.331 \text{ms}^{-1}$ for nitrogen, $0.217 \text{ms}^{-1}$ for helium and $0.133 \text{ms}^{-1}$ for argon, which may be associated with the increasing viscosity of the gases of $1.79 \times 10^{-5}$ Pa s, $1.99 \times 10^{-5}$ Pa s and $2.27 \times 10^{-5}$ Pa s in the same order. The increasing fluid friction between the bulk and surface layers causes a greater dissipation of energy in the trapping phase and results in a lower escape velocity. Consequently, this could indicate that non-linear behaviour is suppressed for the gases with higher viscosities.

According to Figures 3(a), 3(c) and 3(d) which display results for the various degrees of wetting, our model accurately reflects the influence of wetting intrinsically through the probability parameters. Qualitatively, the stronger fluid-solid attraction for a hydrophilic surface should lead us to expect a higher value of $p_s$ and a lower value of $p_e$. All other parameters are constant; this results in a higher value of the coefficient $C_1$. Referring to Figure 3(a), the contact angles for DI Water and OTE-Mica, tetradecane and OTE-Mica, and tetradecane-HAD and Mica are 110°, 44° and 22°, respectively, in the descending order of hydrophobicity. This appears to correspond with the diminishing values of $2.59 \times 10^{-7}$ m, $3.83 \times 10^{-10}$ m and $2.38 \times 10^{-10}$ m obtained for $C_1$. The experimental results of Huang *et al.* (2006) in Figures 3(c) and 3(d) also showed a similar trend with a larger $C_1$ value of $7.16 \times 10^{-22}$ m for the hydrophobic surface compared to $9.23 \times 10^{-27}$ m for the hydrophilic one. The relationship between $C_2$ and viscosity that is apparent in the gaseous slip experiments was also evident in the experiments of Zhu & Granick (2001) in Figure 3(a). Under similar experimental conditions, the tetradecane with a higher viscosity of $2.08 \times 10^{-3}$ Pa s has a lower $C_2$ value of $3.83 \times 10^{-10} \text{ms}^{-1}$ compared to that of water of viscosity $8.9 \times 10^{-4}$ Pa s and the $C_2$ value of $2.59 \times 10^{-7} \text{ms}^{-1}$.

The lack of analytical expressions for the probabilities $p_e$ and $p_m$ confounds the task of obtaining physically sound estimates of their values. One would be tempted to estimate the values of $p_s$, $p_m$ and $p_e$ based on the best-fit coefficients for each data set. However, this requires the approximate values of other parameters such as the free surface diffusion velocity and the friction coefficient for the specific gas-solid pair, which are not readily available in most cases. On the other hand, the sticking probability $p_s$ may be estimated but this too requires the approximation of certain parameters that are elaborated upon below.

In activated adsorption, $p_s$ can be evaluated using the expression

$$p_s = f(\theta)\exp\left(-\frac{E_a}{k_B T}\right) \qquad (41)$$

where $E_a$ is the activation energy and $f(\theta)$ is the surface coverage factor - equivalent to the probability of landing on a vacant site in ideal adsorption. In the case of non-activated adsorption, $p_s$ is a function of $E_n = E_0 \cos^2\theta_{in}$ in what is termed as normal energy scaling if the potential energy surface is only considered in one-dimension along the normal direction. It was determined through the empirical fits of sticking probability data from molecular beam experiments and later theoretically derived that has the sigmodal form (Michelsen & Auerbach 1991, Luntz 2000)

$$p_s = p_{s,\text{sat}}\left[1 + \text{erf}\left(\frac{E_n - E_{n,c}}{W}\right)\right] \qquad (42)$$

where $p_{s,\text{sat}}$ is the saturation sticking probability, $E_{n,c}$ is the value of $E_n$ at the point of inflection on the curve and $W$ is the width of the potential barrier distribution. However, there is now evidence that the sticking probability could scale with the total kinetic energy rather than just the normal energy scaling (Thorman & Bernasek 1981). A possible reason for total energy scaling is the presence of corrugation, which introduces a coupling between the parallel and perpendicular components of velocity. Furthermore, it has been suggested that the prevalence of normal energy scaling could be a fortuitous outcome of the collective effects of energetic and geometric corrugations (Darling & Holloway 1994). The sticking probability in total energy scaling varies as $s = s(E_n, E_t)$ where the parallel energy scaling $E_n = E_0 \sin^2\theta_{in}$. Interestingly, total energy scaling results in higher-order shear rate dependence. Strictly speaking, the sticking probability corresponding to the instantaneous surface coverage should be used. This true value is different from the initial sticking probability prescribed for an adsorbate-free surface. The initial sticking probability is a function of molecular and steric factors that include the incident angle, kinetic energy, temperature, relative orientation of the adsorbate and substrate particles and the location of collision on the substrate. These factors have a strong influence on activated adsorption, which typically exhibits a low initial sticking probability, but not on non-activated adsorption as the initial sticking probability is near unity.

In the literature, the contribution to molecular slip by adsorbed molecules in the mobile state $u_m$ has been suggested to originate from a surface diffusion mechanism of thermally activated surface hops between adjacent adsorption sites (Ruckenstein & Rajora 1983, Yang 2010, Wang & Zhao 2011). The estimates of the slip velocities occurring from this particular mechanism have been shown to be relatively small compared to experimentally measured values (Bowles *et al.* 2011). Therefore, this form of molecular slip accounts for a smaller fraction of the overall slip velocity compared to the contributions by molecules in the other adsorption states. For the slip of liquids on solid surfaces, the mobile adsorbed molecules are expected to make a more significant contribution as the more tightly packed molecular arrangement diminishes the effect of scattering states. Hence, this may imply that the migration of molecules directly across solid surfaces could arise from other surface diffusion mechanisms.